\documentstyle[preprint,aps]{revtex}
\begin{document}

\draft

\title{Localization of charged quantum particles in a static\\
random magnetic field}

\author{A.G. Aronov\cite{AGA}, A.D. Mirlin\cite{SM} and P. W\"olfle}
\address{Institut f\"ur Theorie der Kondensierten Materie,\\
Universit\"at Karlsruhe, 76128 Karlsruhe, Germany}

\date{December 20, 1993}

\maketitle

\begin{abstract}
We consider a charged quantum particle in a random magnetic field with
Gaussian, delta-correlated statistics.  We show that although the single
particle properties are peculiar, two particle quantities
such as the diffusion constant can be calculated
in perturbation theory.  We map the problem onto a non-linear
sigma-model for Q-matrices of unitary symmetry with renormalized
diffusion coefficient for which all states are known to be
localized in $d=2$ dimensions.  Our results compare well with
recent numerical data.
\end{abstract}


\narrowtext

The problem of a charged quantum particle moving in a static
random magnetic field in two dimensions has received renewed interest
recently \cite{Alt,Khv,Pry,Sug,Avi}.  For one, the problem may be
considered as a limiting case of a system of particles interacting
via a gauge field.  Models of this type have been proposed to describe
a state with charge-spin separation of the conduction electrons in
High-T$_c$ superconductors \cite{Iof,Nag}.
Secondly, an experimental realization of a random magnetic field
due to the pinned vortex lines of a superconducting layer on
top of a semiconductor heterostructure has been reported
recently \cite{Geim}.
Thirdly, the problem is thought to be
relevant for the quantum Hall effect in the limit of the half-filled
Landau level \cite{Kal,Hal}.

A number of numerical investigations have been performed, with conflicting
results.  In Ref. \cite{Kal} it was argued on the basis of results
of numerical diagonalization on square lattices of up to 10$^4$ sites
for zero-average random flux per plaquette and in addition
site-diagonal disorder and a uniform magnetic field that localization
can be suppressed by the random flux.  In Ref. \cite{Avi} the
conductance of a square lattice of quantum wires subject to an random
magnetic flux per plaquette, distributed uniformly between $-\phi_0/2$
and $\phi_0/2$ was calculated numerically ($\phi_0$ is the flux
quantum).  Although no definite conclusion could be drawn, the results
were found to be consistent with the existence of extended states
and a mobility edge.  In contrast, the results of applying the finite
size scaling method of MacKinnon and Kramer to the random magnetic
field problem reported in Ref. \cite{Sug} suggested that all states
are localized by a random magnetic field.  Since the localization
length for a two-dimensional disordered system may be very large, it is
obvious that numerical studies of systems of finite size are of limited
value in deciding the principal question whether there exist extended
states in these systems.

In this letter we show that the problem of charged particles in a static
random magnetic field can be mapped onto a nonlinear sigma-model
of unitary matrices.  The latter model has been proposed for disordered
systems containing random spin scattering centers as well as models
featuring random phase fluctuations of the hopping matrix elements
of a tight-binding Hamiltonian \cite{Weg,Ef1}.  Perturbation theory for this
model yields a divergent quantum correction to the conductance in two-loop
order \cite{Brez}.  As a consequence the scaling function in two
dimensions remains negative, leading  to the result that
all states are localized for these models.

We study the transport properties of a charged spinless
quantum particle (mass $m$, charge $e$) in two dimensions in a static
random magnetic field $\vec H(\vec r) = \vec\nabla \times
\vec A(\vec r)$ normal to the plane, as defined by the Hamiltonian
\begin{equation}
H = {p^2\over 2m} - {e\over mc} \vec p \cdot \vec A + {e^2\over 2mc^2}
{\vec A}\:^2 ,
\label{1}
\end{equation}
where $\vec p = - i\vec\nabla$ is the momentum operator and the
Coulomb gauge $(\vec\nabla\cdot\vec A = 0)$ has been used ($\hbar = 1$).
The magnetic field is assumed to be Gaussian distributed and delta-correlated,
with vanishing mean and variance in Fourier space
\begin{equation}
\langle H (\vec q) H (-\vec q) \rangle = \langle h^2\rangle
\label{2}
\end{equation}
Accordingly, the variance of the vector potential is $\langle A_\alpha
(\vec q) A_\beta (-\vec q)\rangle = (1/q^2) \langle h^2\rangle$
$\delta_{\alpha\beta}^T
(\hat q)$ where $\delta_{\alpha\beta}^T (\hat q) =
\delta_{\alpha\beta} - \hat q_\alpha\hat q_\beta$ accounts for the
transverse character of the field $(\hat q = \vec q/\mid\vec q\mid)$.
By contrast, an independent random distribution of phases of the
hopping matrix elements correspons to a delta-correlated distribution
of vector potentials given by
$\langle A_\alpha (\vec q)A_\beta (-\vec q)\rangle
= \langle a^2\rangle \delta_{\alpha\beta}^T(\hat q)$.  We will comment
on this case later.

Assuming the fluctuation strength of the magnetic field,
$\langle h^2\rangle$, to be
weak, we first consider perturbation theory.  The standard Feynman diagram
language for impurity scattering may be employed, with the impurity
line describing scattering of a particle from momentum state
$\mid\vec p + {\vec q/ 2}\rangle$  into state $\mid\vec p{\;}' +
\vec q/2 \rangle$ and a hole from
$\mid \vec p - \vec q/2\rangle$ into $\mid \vec p{\;}'
- \vec q/2\rangle$ being given by
\begin{equation}
w_{\vec p\vec p{\;}'}(q) = v_0^2 k^{-2}\:\Big\{ (\vec p + \vec p{\;}')^2
- q^2 - [(\vec p + \vec p{\;}')\cdot \hat k]^2 + (\vec q \cdot\hat k)^2
\Big\}
\label{3}
\end{equation}
where $\vec k = \vec p - \vec p{\;}'$ is the transferred momentum
and $v_0^2 = e^2 \langle h^2\rangle/4m^2c^2$ is a
velocity squared characteristic of the strength of disorder.
Note that $w_{\vec p\vec p{\;}'}$ is strongly singular in the forward
direction $(\vec p = \vec p{\;}')$, due to the long range of the vector
potential fluctuations, even though the magnetic field fluctuations
are assumed to be short-ranged.  In lowest order the imaginary part of the
(advanced)
single particle self energy $\Sigma^A(\vec p,E)$ on the energy
shell $(E = p^2/2m )$ is given by
\begin{eqnarray}
{\rm Im} \Sigma^A \equiv {1\over 2\tau}  && =
\pi\int (dp{\;}') w_{pp{\;}'}(0) \delta({p^2\over 2m} - {p{\;}'^2\over
2m})\nonumber \\
&& = \pi N_0 v_0^2 \int_0^{2\pi} {d\phi\over 2\pi} \cot^2 {\phi\over 2}
\label{4}
\end{eqnarray}
Here  $(dp) = d^2p/(2\pi)^2$,
 $\hat p\cdot \hat p{\;}' = \cos \phi$ and $N_0 = {m\over 2\pi}$
is the density of states.
The $\phi$ integral in (\ref{4}) is strongly divergent at $\phi = 0$, which
may be traced to the contribution of vector potential fluctuations
in the limit $q\rightarrow 0$.  A self-consistent treatment
of the divergence leads to a much weaker dependence as discussed
below after (\ref{17}).  Nonetheless, the single particle
relaxation rate $1/2\tau$ can be expected to diverge, which might
lead to the generation of a branch cut in the single particle Green's
function, as argued in Ref. \cite{Alt,Khv} (see however \cite{note}).  We
will show below that the
transport relaxation rate and the diffusion constant are nonsingular,
in agreement with results of a naive perturbation theory \cite{Khv}.
This may be done by regularizing the infrared divergence in (\ref{4})
in a convenient way.  We will use a soft cut-off, replacing $\cot^2 {\phi
\over 2}$ in (\ref{4}) by $w(\phi) =
\cos^2 {\phi\over 2}/(\sin^2 {\phi\over 2} +
\gamma^2)$, although the precise form of the cutoff is not
important.  At the end of the calculation we take the limit $\gamma
\rightarrow 0$.

Next, let us consider the diffusion propagator $\Gamma$ (``diffuson''),
obtained by summing the particle-hole ladder diagrams
\begin{equation}
\Gamma_{\vec p\vec p{\;}'}(\vec q,\omega) = w_{\vec p\vec p{\;}'}(\vec q)
 +\!\! \int (dp^{''})\: w_{\vec p\vec p^{{\;}{''}}}(\vec q)
G_{\vec p_+^{{\;}{''}}}^R
(E_+)G_{\vec p_-^{{\;}{''}}}^A(E_-)\Gamma_{\vec p^{{\;}{''}}\vec p{\;}'}
(\vec q,\omega)
\label{5}
\end{equation}
where
\begin{equation}
G_{\vec p}^{R,A}(E) = \Big[E - {p^2\over 2m} \pm {i\over 2\tau}\Big]^{-1}
\label{a}
\end{equation}
and $\vec p_\pm^{{\;}{''}} = \vec p^{{\;}{''}} \pm \vec q/2$, $E_\pm
= E \pm \omega/2$.  The solution of (\ref{5}) may be easily obtained
in terms of the eigenfunction of the operator $w_{\vec p\vec p{\;}'}(q = 0)$
(see Ref. \cite{Bhatt}).  In two dimensions and for $\mid\vec p\mid = \mid\vec
p{\;}'
\mid$ one has
\begin{equation}
w_{\vec p\vec p{\;}'}(\vec q = 0) = \sum_n w_n \chi_n^*(\hat p)
\chi_n(\hat p{\;}')
\label{6}
\end{equation}
with
\begin{equation}
w_n = v_0^2 \int_0^{2\pi} {d\phi\over 2\pi} w(\phi)
e^{-in\phi}.
\label{b}
\end{equation}
and
\begin{equation}
\chi_n (\hat p) = e^{in\phi}
\label{c}
\end{equation}
Here $\phi$ is the polar angle of $\hat p$.  The leading contribution
to $\Gamma_{\vec p\vec p{\;}'}$ is obtained as
\begin{equation}
\Gamma_{\vec p\vec p{\;}'}(q,\omega) = {1\over -i\omega + Dq^2}
{1\over 2\pi N_0\tau^2}
\label{7}
\end{equation}
where the diffusion coefficient $D$ is given by
\begin{equation}
D = {1\over 2} v^2\tau (1 + {w_1\over w_0 - w_1}) = {1\over 2}
v^2\tau_{tr}
\label{8}
\end{equation}
and $v = p/m$.
Here we have defined the transport relaxation time $\tau_{tr}$
as
\begin{equation}
{1\over\tau_{tr}} = mv_0^2 \int {d\phi\over 2\pi}\cot^2 {\phi\over 2}
(1- \cos\phi) = mv_0^2,
\label{9}
\end{equation}
In contrast to $\tau$, the  transport time $\tau_{tr}$
is finite in the limit $\gamma \rightarrow 0$.

In a time reversal invariant system, the coherent backscattering
described by a diffusion pole in the particle-particle ladder
diagrams, the socalled Cooperon, plays a dominant role.  The Cooperon
$C_{\vec p\vec p{\;}'}(\vec q,\omega)$ obeys the integral equation
(\ref{5}), with $w_{\vec p\vec p{\;}'}(\vec q)$ replaced  by
$w_{\vec k\vec k{\;}'}(\vec Q)$, where $\vec k = {1\over 2}
(\vec p - \vec p{\;}' + \vec q)$,
$\vec k{\;}' = {1\over 2}(\vec p{\;}' - \vec p + \vec q)$ and
$\vec Q = \vec p + \vec p{\;}'$.  The fact that the vector potential
$\vec A$ couples to the particle momentum $\vec p$ (see (\ref{1})),
which changes the sign under time reversal, leads to the relation
$w_{\vec k \vec k{\;}'} (\vec Q) = - w_{\vec p\vec p{\;}'}(\vec q)$.
Correspondingly, the Cooperon is finite in the limit $\vec q,
\omega\rightarrow 0$ and cannot play any role in bringing about
localization in the present case.

In the following we map the problem onto a nonlinear sigma-model
of unitary symmetry.  The generating functional for two-particle
Green's function of the retarded-advanced type (RA) may be represented
in terms of a functional integral over a
supersymmetric field $\psi = (\varphi_1,
\chi_1,\varphi_2,\chi_2)$,
where $\varphi_{1,2}$ are bosonic and $\chi_{1,2}$ are fermionic
components as \cite{Ef}
\begin{equation}
Z = \int D[\psi] \exp - (S_0 + S_1)
\label{10}
\end{equation}
where
\begin{equation}
S_0 = -i\int d^2r\Big\{\bar\psi \Lambda (E + {1\over 2m}
\nabla^2)\psi - i {\omega\over 2}(\bar\psi\psi)\Big\}
\label{d}
\end{equation}
and
\begin{equation}
S_1 = - i {e\over mc} \int d^2r \Big\{\bar\psi
\Lambda (- i \vec A \cdot\vec\nabla)\psi\Big\}.
\label{e}
\end{equation}
 The 4 x 4 matrix
$\Lambda$ is diagonal, $\Lambda = {\rm diag} (1,1, -1, -1)$.

Averaging over the vector potential one finds that $S_1$ in (15)
has to be replaced by
\begin{equation}
S_1^{eff} = - i \sum_{\vec k,\vec k{\;}'\atop q < q_0}
\Lambda_\alpha \bar\psi_{\vec k + \vec q}^\alpha \psi_{\vec k}^\beta
w_{\vec k\vec k{\;}'}\bar\psi_{\vec k{\;}' - \vec q}^\beta
\psi_{\vec k{\;}'}^\alpha \Lambda_\beta
\label{11}
\end{equation}
where $w_{\vec k,\vec k{\;}'}$ is defined in (\ref{3}), and only
long-wave length fluctuations are considered $(q < q_0)$. It is useful
to introduce the representation of $w_{\vec k,\vec k{\;}'}$ in terms
of eigenfunctions (\ref{6}),
and to define ``density'' fields \cite{Bhatt}
\begin{equation}
\rho_{n,\vec q}^{\alpha\beta} = \sum_{\vec k}\Lambda_\alpha
\bar\psi_{\vec k + \vec q}^\alpha \psi_{\vec k}^\beta \chi_n (\hat k)
\label{12}
\end{equation}
in terms of which $S_1^{eff}$ can be written as
\begin{equation}
S_1^{eff} = - i \int d^2r \sum_n w_n \bar\rho_n^{\alpha\beta}(\vec r)
\rho_n^{\beta\alpha}(\vec r).
\label{13}
\end{equation}
As usual, the interaction term may be decoupled with the aid of
Hubbard-Stratonovich fields $Q_n(r)$, which are (4 x 4)
supermatrices of unitary symmetry.  The functional integration
over the primary fields may be performed, yielding
\begin{equation}
Z = \int D[Q] \exp - \tilde S\{ Q \}
\label{14}
\end{equation}
where the effective action of the $Q$-fields is at first given by
\begin{equation}
\tilde S = \int d^2r \Big\{{\rm Str}\;\; \ln\;\; G^{-1}
- {1\over 2} \sum_n w_n\;\;{\rm Str}\;\; Q_n^2\Big\}
\label{15}
\end{equation}
where Str denotes the supertrace and
\begin{equation}
G_p^{-1} = \hat\epsilon - {p^2\over 2m}  + i \sum_n
w_n \chi_n(\hat p)Q_n\Lambda
\label{f}
\end{equation}
and $\hat\epsilon = {\rm diag}\;\; (E + {\omega\over 2}, E + {\omega\over 2},
E - {\omega\over 2}, E - {\omega\over 2})$.  The saddle point of
$\exp (- \tilde S)$
is at $Q_n = Q_n^{(0)}$, where $Q_n^{(0)}$ is
\begin{eqnarray}
 Q_n^{(0)} && = i \int (dp) \chi_n(\hat p) G(p)\Lambda\nonumber \\
&& = i\delta_{n0}  \int (dp)\Big[\hat\epsilon - {p^2\over 2m}
 + iw_0 Q_0^{(0)}\Lambda\Big]^{-1}\Lambda
\label{16}
\end{eqnarray}
so that the Green's function at the saddle point is given by
\begin{equation}
G(p) = \Big[ \hat\epsilon - {p^2\over 2m} + {i\over 2\tau}
\Lambda\Big]^{-1}
\label{17}
\end{equation}
Equations (\ref{16}) and (\ref{17}) are a statement of the self-consistent
Born approximation for the single-particle relaxation rate ${1\over\tau}$.
In contrast to the lowest order expression (\ref{4}) for ${1\over \tau}$,
which scales with the cutoff $\gamma$ as $\tau^{-1} \propto \gamma^{-1}$,
the self-consistent value is given by $\tau^{-1} = [{4\over\pi}p^2
v_0^2\;\ln\,(1/E\tau\gamma)]^{1/2}$ and hence shows a weaker
divergence as $\gamma\rightarrow 0$ \cite{note}.

We now expand the action around the saddle point:
\begin{eqnarray}
\tilde S = \tilde S_0  + {1\over 2}\int (dq) \Big\{&& \int (dp)
\sum_{n,m}w_n\chi_n (\hat p)w_m \chi_m (\vec p + \vec q)
\;{\rm Str}\:\Big[G(\vec p)\Lambda \delta Q_n(\vec q)G(\vec p + \vec q)
\Lambda \delta Q_m(-\vec q)\Big] \nonumber\\&& - \sum_n
w_n\;{\rm Str}\:\Big[\delta Q_n(q)\delta Q_n(-q)\Big]\Big\}
\label{f1}
\end{eqnarray}
The $p$-integral is only finite for the products $G^RG^A$, which are
generated by the off-diagonal components of $\delta Q$, denoted $\delta
\tilde Q$.  In the limit of small $\vec q,\omega$, using
\begin{eqnarray}
&&\int (dp) \chi_n(\hat p)\chi_m (\hat p)  G^A(E - {\omega\over 2},
\vec p)G^R (E + {\omega\over 2}, \vec p + \vec q)\nonumber \\
&& = 2\pi N_0\tau\Big[\delta_{nm} + i\tau \Big(\omega\delta_{nm} -
{pq\over m}b_{nm}^{(1)}\Big) - \Big({pq\tau\over m}\Big)^2 b_{nm}^{(2)}\Big],
\label{g}
\end{eqnarray}
where
$b_{nm}^{(\ell)} = \langle\chi_n\chi_m \cos^\ell \phi\rangle$, $\ell = 1,2$,
we find
\begin{eqnarray}
\tilde S  = \tilde S_0 - {1\over 2}\int (dq)
\Big\{&& \sum_n w_n (1- {w_n\over w_0})\;{\rm Str}\:[\delta \tilde Q_n(q)
\delta \tilde Q_n(-q)]\nonumber \\
&& + w_0\tau (-i\omega + D_0q^2)\;{\rm Str}\:[\delta\tilde Q_0
(q)\delta\tilde Q_0(-q)]\nonumber \\
&& + {i\over 2}w_1\tau ({pq\over m})\sum_{n = \pm 1}\;{\rm Str}\:
[\delta\tilde Q_0(q) \delta\tilde Q_n(-q)]\Big\}
\label{18}
\end{eqnarray}
As expected, the $n=0$ mode is massless and describes interacting
diffusons.  The coefficient of th $q^2$-term is the bare diffusion
constant, $D_0 = {1\over 2} v^2\tau$, which tends to zero if the
infrared cutoff is taken to zero.  However, the bare diffusion
constant gets dressed by the coupling to the massive $n = \pm 1$ modes.
Indeed, integrating out $\delta Q_1$ and $\delta Q_{-1}$ produces
a renormalization term
\begin{equation}
\Delta\tilde S = - {1\over 2} \int (dq)  D_0\tau {w_0 w_1\over
w_0 - w_1} q^2 \;{\rm Str}\:[\delta\tilde Q_0(q)\delta\tilde Q_0
(-q)]
\end{equation}
which combined with the bare diffusion term has the effect of changing
the bare diffusion constant
$D_0$ into the renormalized $D$ as defined in (\ref{9}).  The final result for
the
effective action is
\begin{equation}
\tilde S  = \tilde S_0 - \pi N_0\int (dq)
\Big\{(Dq^2 - i\omega)\;{\rm Str}\:[\delta\tilde Q_0(q)\delta
\tilde Q_0(-q)]\Big\}
\label{20}
\end{equation}
We note in passing that the coefficients of terms with higher
spatial derivatives of $Q$ will be renormalized in a similar
way and may be expected to be finite as well.

The expansion (\ref{20}) of the action in terms of $\delta\tilde Q_0$
serves to determine the coefficients of the two terms in the nonlinear
sigma model obtained from (\ref{15}) by keeping only the integration
over the saddle point manifold:
\begin{equation}
S_\sigma = {\pi N_0\over 4}\int\! d^2r \Big\{-D\:{\rm Str}\,(\vec\nabla Q \cdot
\vec\nabla Q) - 2i\omega\:{\rm Str}\, (\Lambda Q)\Big\}
\label{h}
\end{equation}
where the rescaled field $Q(\vec r)$ is constrained by $Q^2(\vec r) = 1$.

We have thus shown that the problem of a charged quantum particle in a
static random magnetic field is equivalent to a nonlinear $\sigma$-model
of interacting $Q$-matrices with unitary symmetry.  This model
has been studied extensively \cite{Weg,Ef1,Brez}.
It is known that the Gell-Mann-Low
$\beta$-function describing the scaling behavior of the dimensionless
conductance $g$ (in units of $e^2/h$) with the length of the sample is given
inleading order for large $g$ b
\begin{equation}
{d\,\ln\,g\over d\,\ln\,L} = \beta(g) =
- {1\over 2\pi^2 g^2} + O\Big({1\over g^4}\Big)
\label{x}
\end{equation}
It follows then that all states are localized, and that the
localization length in the weak disorder regime  is given by
\begin{equation}
\xi = \xi_0 \exp  (\pi^2 g_0^2)
\label{y}
\end{equation}
where $g_0 = mv^2 \tau_{tr}/2 = (v^2/4v_0^2)$ is the Drude conductance, and
$\xi_0
\simeq v\tau_{tr}$.
For the case of a delta-correlated distribution of vector potentials
the single particle relaxation rate ${1\over 2\tau}$ does not show
an infrared
divergence, and the nonlinear sigma model may be derived in the
usual way.  The more complete derivation given here leads
to a renormalization of the diffusion constant as $D = 2D_0
= (4\pi N_0 e^2 \langle a^2\rangle/c^2)^{-1}$
where $\langle a^2\rangle$ was defined after (2).

Our results are in good agreement with the available numerical data
\cite{Sug,Avi}.  The authors of these papers studied the lattice
version of  the problem with the maximum possible disorder
corresponding to $v_0 \sim v$.  Accordingly, the typical values of the
``bare'' conductance $g_0$ are of order of unity.  However, when
approaching the center of the band (i.e. when $g_0$ increases),
the localization length was found to grow exponentially \cite{Sug},
in agreement with  (31).  The finite size scaling analysis \cite{Avi}
yields for the localization length $\Lambda$ of a quasi 1D strip
of  width $M$:
\begin{equation}
\Lambda(M) = M f (\xi/M)
\label{32}
\end{equation}
Calculating the scaling function $f(x)$ by using (31) and comparing with
the known result for the localization length of a quasi 1D
system \cite{Ef2}, we obtain \cite{Ref}:
\begin{equation}
f(x) \simeq \cases{{2\over \pi} \sqrt{\ln\;\; x}, & $x \gg 1$\cr
x, & $x \ll 1$\cr}
\label{33}
\end{equation}
This agrees well with the asymptotic behavior of $f(x)$ as
obtained by numerical means in \cite{Sug}, for both $x \gg 1$ and
$x \ll 1$ (see Fig.1). Scaling behavior of the conductance $g(L/\xi)$
obtained in \cite{Avi} (see Fig.4 of Ref.\cite{Avi}) is also compatible
with the scaling law $g(L)\sim(1/\pi)\sqrt{\ln (\xi/L)}$ ($\xi\gg L$)
 which follows from Eqs.(\ref{x}), (\ref{y}).

To summarize, we have shown here that the nonlinear sigma model
description of disordered systems with broken time reversal
invariance holds true even in the case of long-ranged fluctuations
of the vector potential when the single particle properties
are dominated by infrared divergencies.  This is true provided
the magnetic field fluctuations are short ranged.  In the opposite
case of long-ranged magnetic field fluctuations it is conceivable
that  even the transport relaxation rate diverges, signalling
a different physical regime.  One may speculate that then the
topological excitations governing the behavior in the Quantum-Hall-Effect,
where the average magnetic field is finite and large, will play a role.
However, for short-ranged magnetic field fluctuations the topological term
is absent.  Finally, we emphasize that our analysis is restricted
to quenched random magnetic fields.  To what extent this model applies
to dynamical gauge field models remains to be seen.

This work was supported by the Sonderforschungsbereich 195 der
Deutschen Forschungsgemeinschaft (A.G.A.) and by the Alexander
von Humboldt Foundation (A.D.M.).  We thank Frau Rose Schrempp
for her kind help in the preparation of the manuscript.

\section*{Figure caption}

Fig.1. The scaling function $f(x)$, Eq.(\ref{32}),
as obtained by numerical study of the problem in Ref.\cite{Sug}. The
dashed and dot--dashed lines represent the asymptotical behavior for
$x\ll 1$ and $x\gg 1$ respectively, given by Eq.(\ref{33}).

\begin{references}
\bibitem{Ref} We used the definition of the localization length
adopted in Ref. \cite{Sug}, which differs by a factor $2$ from the one used
in Ref. \cite{Ef2}.
\end{references}
\end{document}